\documentclass[prd,twocolumn,showpcs,color,amsmath,amssymb,nofootinbib,preprintnumbers,balancelastpage,english]{revtex4-1}
\usepackage[latin9]{inputenc}
\usepackage{cancel}
\usepackage{graphicx,color}

\makeatletter

\@ifundefined{showcaptionsetup}{}{
 \PassOptionsToPackage{caption=false}{subfig}}
\usepackage{subfig}
\makeatother

\usepackage{babel}
\begin{document}

\preprint{UCI-HEP-TR-2016-10}

\title{Mono-jet Signatures of Gluphilic Scalar Dark Matter}

\author{Rohini M. Godbole$^{1}$, Gaurav Mendiratta$^{1,2}$, Ambresh Shivaji$^{3,4}$ Tim M.P. Tait$^{5}$\\
$^{1}$CHEP, Indian Institute of Science, Bangalore, India 560012\\
$^{2}$IBL, Salk Institute of Biological Studies, 10010 North Torrey Pines Rd, La Jolla, CA, USA 92037\\ 
$^{3}$INFN, Sezione di Pavia, Via A. Bassi 6, 27100 Pavia, Italy\\
$^{4}$Centre for Cosmology, Particle Physics and Phenomenology (CP3), Universit\'{e} Catholique de Louvain, B-1348 Louvain-la-Neuve, Belgium\\
$^{5}$Department of Physics and Astronomy, University of California,
Irvine, CA 92697\\}

\begin{abstract}
A gluphilic scalar dark matter (GSDM) model has recently been proposed
as an interesting vision for WIMP dark matter communicating dominantly with the Standard Model
via gluons.  We discuss the collider signature of a hard jet recoiling against missing momentum (``mono-jet")
in such a construction, whose leading contribution is at one-loop.
We compare the full one-loop computation with an effective field theory (EFT) treatment, and
find (as expected) that EFT does not accurately describe regions of parameter space 
where mass of the colored mediator particles are comparable to the experimental
cuts on the missing energy.  We determine bounds (for several choices of SU(3)
representation of the mediator) from the $\sqrt{s}=$ 8 TeV data,
and show the expected reach of the $\sqrt{s}=$ 13 TeV LHC and a future 100 TeV
$pp$ collider to constrain or discover GSDM models.
\end{abstract}


\maketitle

\section{Introduction}
\label{sec:intro}

An overwhelming body of evidence from astrophysical observations points
toward a large, invisible component of the matter content of the universe, usually
termed dark matter (DM).  Any candidate for DM has to be stable
over the lifetime of the universe, requiring either extremely small couplings
with the Standard Model (SM) particles, or the presence of additional symmetries forbidding its decay.
If the DM has sufficiently strong coupling to SM particles, it can be probed at high energy hadron colliders,
where it typically escapes from the detectors and thus appears as an imbalance in the visible momentum.
Not surprisingly, ``missing transverse energy" (MET, or $\cancel{E}_{T}$)
channels are an important part of the physics menu at the
Large Hadron Collider (LHC) \cite{Morrissey:2009tf}.

In many models of dark matter, including some of the more popular theories such
as supersymmetric extensions of the SM,
the processes giving rise to MET occur at tree level.  Such processes include
production of one or more mediator particles which decay into dark matter and/or
visible radiation.  In the limit in which the mediator particles are heavy, all theories
flow into a universal effective field theory (EFT) consisting of non-renormalizable
interactions between the dark matter and the 
SM \cite{Beltran:2008xg,Cao:2009uw,Beltran:2010ww,Goodman:2010yf,Bai:2010hh,Goodman:2010ku}.
When the mediator particles are light, and must be explicitly included in the description,
these are supplanted by simplified models (see e.g. \cite{Abercrombie:2015wmb,Abdallah:2015ter}).

However, another interesting class of theories has the dark matter communicating with the SM primarily through loop
processes.  This further opens new portals of interaction, such as communication with SM gluons, whose tree level
interactions are strongly constrained by gauge invariance.The particle in the loop may be an SM particle, for example see \cite{Mattelaer:2015haa}, or a BSM particle. In \cite{Godbole:2015gma}, such a model was explored
in which the dark matter ($\chi$) is a scalar particle 
whose primary renormalizable interaction is through a quartic interaction with an exotic colored scalar
($\phi$), leading to a one-loop coupling of two dark matter particles to gluons (see also Ref.~\cite{Bai:2015swa}
for related discussion).
While simple, such a construction leads to novel features.  For example, the $Z_2$ symmetry posited to
insure that $\chi$ is stable need not act on the $\phi$, which can decay into hadronic jets.
Such particles can look somewhat like the squarks of an $R$-parity violating supersymmetric model
(despite the current framework being a model of dark matter), and are rather weakly constrained
by LHC searches for resonant structure in 
dijet and 4-jet final states
\cite{Han:2010rf,Chivukula:2015zma,Khachatryan:2015dcf,Khachatryan:2014lpa,ATLAS:2015nsi}
(see \cite{Godbole:2015gma} for detailed discussion).
However, it is the mono-jet process, in which a pair of dark matter recoils against a hadronic jet,
which directly probes the DM and its coupling to the SM.  Such a theory is a natural 
theoretical laboratory to explore the features of a construction in which the dark matter 
couples to gluons through loops.

Ref. \cite{Godbole:2015gma} confined its discussion of the mono-jet signal to 
the limit of heavy mediators, in which the loop diagram matches on to the operator
C5 in the EFT \cite{Goodman:2010ku}.  In the current work, we extend this result to the
case of lighter mediators by performing the full one-loop calculation of the mono-jet process,
valid for all mediator masses.  Our primary aim is to understand the current and future
limits from the LHC and a future 100 TeV $pp$ collider (such as the proposed future circular collider (FCC))
on this interesting theory of dark matter, but as a by-product we also examine the
systematics of how the full theory transitions into the EFT description in the heavy mass limit.

This article is organized as follows.  In section \ref{sec:Effective-field-theory}, 
we review the most important features of the gluphilic scalar dark matter (GSDM) model,
and its mapping into the EFT language at large mediator masses.
In section \ref{sec:Monojet-loop-induced-cross},
we detail the loop calculation for the mono-jet process, discuss the role of 
$gg$, $gq$, and $q\bar{q}$ initial states, and compare the results derived from the EFT limit. 
In section \ref{sec:Results-and-projections},
we show the bounds interpreted from current LHC 8 TeV, run-II data
and the projected mono-jet cross-sections along with the expected leading
order backgrounds at the LHC Run II (13 TeV) and the FCC (100 TeV). In section
\ref{sec:Summary}, we conclude with some outlook.

\section{GDSM Model and Mapping to EFT}
\label{sec:Effective-field-theory}

The basic module for a scalar dark matter (taken to be complex here, though the generalization
to a real scalar is trivial) coupling to a colored scalar $\phi$ consists of renormalizable interactions:
\begin{equation}
\lambda_{sh}~H^{\dagger}H\phi^{\dagger}\phi + \lambda_{d}~\chi^{*}\chi\phi^{\dagger}\phi,
\label{eq:potential}
\end{equation}
where $H$ is the SM Higgs doublet.  The quartic $\lambda_{sh}$ implements DM coupling
to the SM Higgs, and is known as the Higgs portal \cite{Burgess:2000yq}.  It leads to interesting
predictions for Higgs physics which generically rule out regions of the parameter space consistent
with the dark matter being a thermal relic \cite{Djouadi:2012zc}.
We assume this coupling is small enough to provide subdominant effects in our analyses below.
In that limit, the parameters are the coupling $\lambda_d$,
the masses of $\phi$ and $\chi$, and the choice of
SU(3)$_{C}$ representation $r$ of $\phi$. 
Generically it is desirable to include couplings between the mediators and the SM quarks so that they can decay.
Such couplings depend on $r$, and are thus more model-dependent.  We will assume that they
are small enough so as to play little role in dark matter production at colliders.

When $m_{\phi}$ is large compared to the energies of interest, $\phi$ can be integrated out, leaving
behind the non-renormalizable contact interaction C5:
\begin{equation}
{\cal L}_{\rm EFT} =
\frac{\lambda_d \alpha_s T_r}{48 \pi}  \frac{1~}{m_{\phi}^2} 
~ |\chi|^2 G^a_{\mu \nu} G^{a \mu \nu}~,
\label{eq:eftmatching}
\end{equation}
where $G^a_{\mu \nu}$ is the gluon field strength and
$T_r$ is the Casimir corresponding to the choice of the representation of $\phi$ under SU(3).
In \cite{Godbole:2015gma}, the bound on this operator from the CMS 
mono-jet analysis with $\cancel{E}_{T} > 500$ GeV \cite{Khachatryan:2014rra}
was derived to be,
\begin{equation}
\frac{\lambda_{d}T_{r}}{48\pi}\frac{1}{m_\phi^{2}}\le\frac{1}{\left(207\mbox{ GeV}\right)^{2}}.
\end{equation}
Due to the large cut on $\cancel{E}_{T}$, the EFT description is only expected to
be a good approximation when $m_\phi \gtrsim$~TeV.  Thus, the bound really only applies
self-consistently for very large $\lambda_{d}$ and/or $T_{r}$.
A meaningful estimation of the bound really requires the full calculation in the simplified model
framework.

\section{one-loop Mono-jet Rates}
\label{sec:Monojet-loop-induced-cross}

\begin{figure*}[!t]
\hfill{}\subfloat[qq]{\includegraphics[scale=0.65]{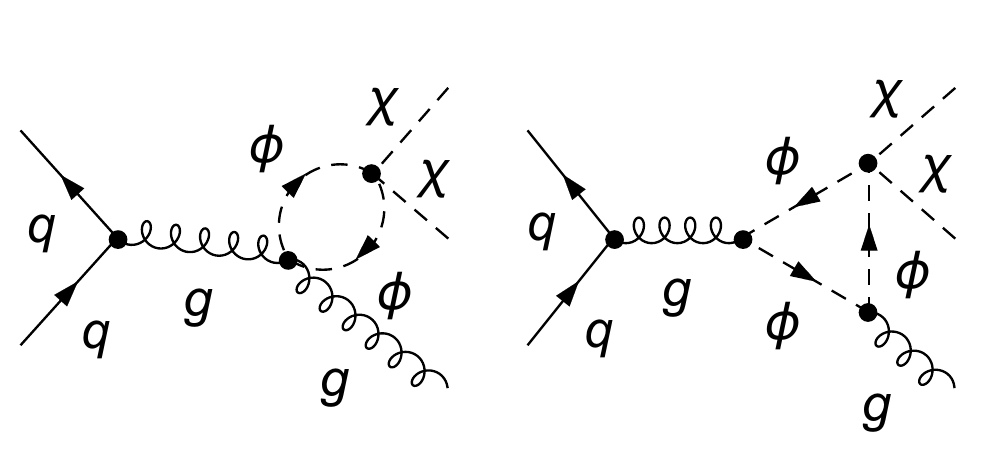}

}\hfill{}\subfloat[qg]{\includegraphics[scale=0.65]{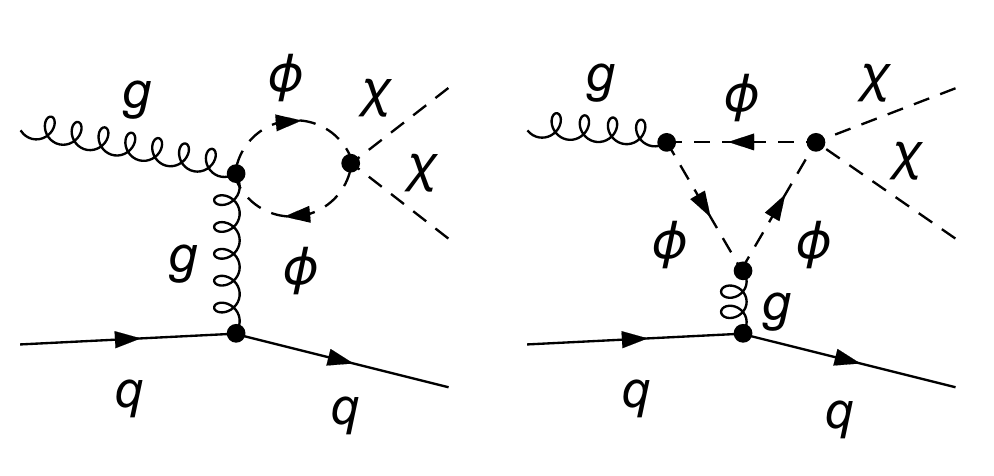}

}\hfill{}

\hfill{}\subfloat[gg]{\includegraphics[scale=0.65]{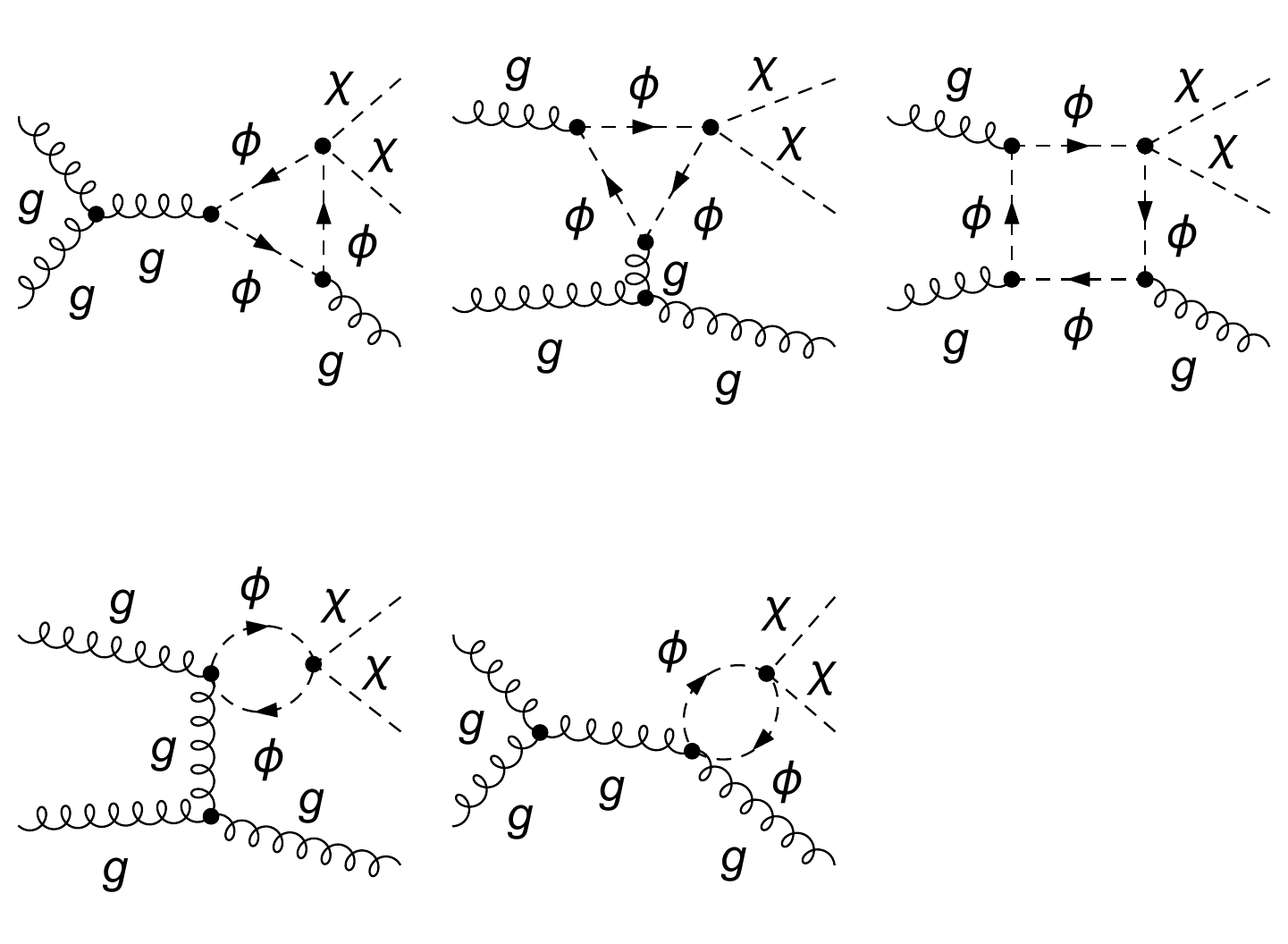}

}\hfill{}

\caption{\label{fig:Representative-diagrams}Representative diagrams for the
subprocesses contributing to $pp\to j \chi \chi^*$ at a hadron collider.}
\end{figure*}

The mono-jet process, $p p \to j \chi \chi^*$,
receives contributions at the parton level from processes:
\begin{eqnarray}
g g & \rightarrow & \chi ~\chi^* ~g, \nonumber \\
g q(\bar{q}) & \rightarrow &  \chi ~\chi^* ~q(\bar{q}), \nonumber \\
q \bar{q}  & \rightarrow & \chi~ \chi^* ~g,
\end{eqnarray}
where $q$ denotes any light quark, $q=u,d,s,c$. The one-loop $\phi$-mediated 
diagrams contributing to $gg$ channel are box, triangle and bubble type, while those 
contributing to $q\bar{q}$ and $gq$ subprocesses are triangle and bubble type.
Representative one-loop Feynman diagrams for all three initial states are shown in 
Figure \ref{fig:Representative-diagrams}.  
Due to the charge conjugation symmetry, the triangle diagram with $gg\phi^\dagger\phi$ coupling does not contribute 
in $gg$ subprocess, and so is not displayed in the Figure \ref{fig:Representative-diagrams}.

\subsection{ Method of Calculation \label{sec:Calculation}}

We evaluate the cross section for $pp\to j\chi\chi^*$ using three different techniques: 
\begin{enumerate}
\item In the first method we compute the one-loop amplitude for each
subprocess using an in-house code based on the tensor reduction method suggested
by Oldenbourgh and Vermaseren ({\tt OV}) \cite{vanOldenborgh:1989wn,Shivaji:2013cca}
with dimensional regularization ($d=4-2\epsilon$). We implement the reduction
routines in Fortran, and evaluate all of the required scalar
integrals using the OneLOop library \cite{vanHameren:2010cp}. 
The reduction routines are interfaced with {\tt RAMBO}~\cite{Kleiss:1985gy}, a Monte Carlo multiparticle phase space 
generator, to compute cross-sections and desired kinematic distributions. We have ensured 
the internal consistency of the calculation by checking the ultraviolet finiteness of the 
one-loop amplitudes and their gauge invariance with respect to the external gluons and currents.

\item To cross check, we calculate the same amplitudes using publicly available standard packages. We use FeynRules \cite{Christensen:2008py} to process an implementation of the GDSM Lagrangian, and pass the output to
the FeynArts package \cite{Hahn:2000kx}. We generate the relevant diagrams and corresponding amplitudes in FeynArts and use FormCalc \cite{Hahn:2006zy} to
perform the Passarino-Veltman ( PV) reduction \cite{Passarino:1978jh} and use the LoopTools library \cite{Hahn:1998yk} to 
evaluate scalar integrals numerically. We find that for a given phase space point the 
one-loop amplitudes computed using the two methods are in excellent agreement.

\item
We also compute the monojet cross section utilizing Madgraph (MG5) \cite{Alwall:2014hca}, whose recently added
NLOCT \cite{Degrande:2014vpa} framework can handle loop calculations.  The advantage of using this framework is that many other possible signatures of the model can 
be studied in an automated way. Moreover, jet matching, parton showering, and detector simulations
can also be relatively easily included, if desired.
In the results obtained here, we found that with small $p_{T}$ cuts, the MG5 produced cross-sections are consistent with those obtained using above methods to within $\sim1\%$.  However, at large $p_{T}$ the MG5 generated cross sections are subject to large Monte Carlo fluctuations and the method requires generation of a very large number of simulated events to arrive at a stable answer.  For this reason, we employ the results obtained using the in-house code in the first method listed above in the remainder of this work.
\end{enumerate}

\begin{figure*}[!t]
\centering{}\subfloat[$m_{\phi}=10$ GeV]{\begin{centering}
\includegraphics[scale=0.28]{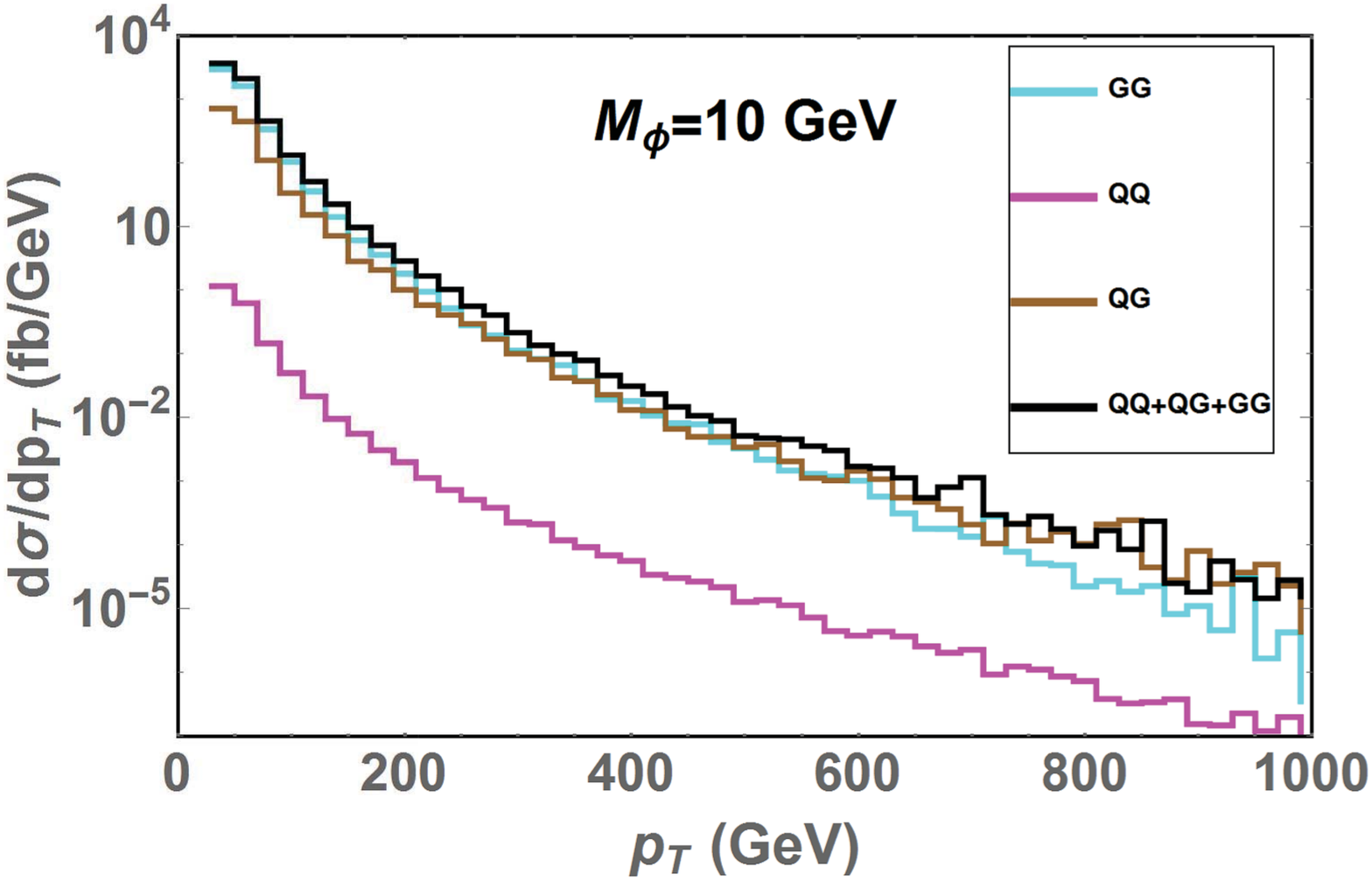}
\par\end{centering}

}\hfill{}
\subfloat[$m_{\phi}=100$ GeV]{\begin{centering}
\includegraphics[scale=0.28]{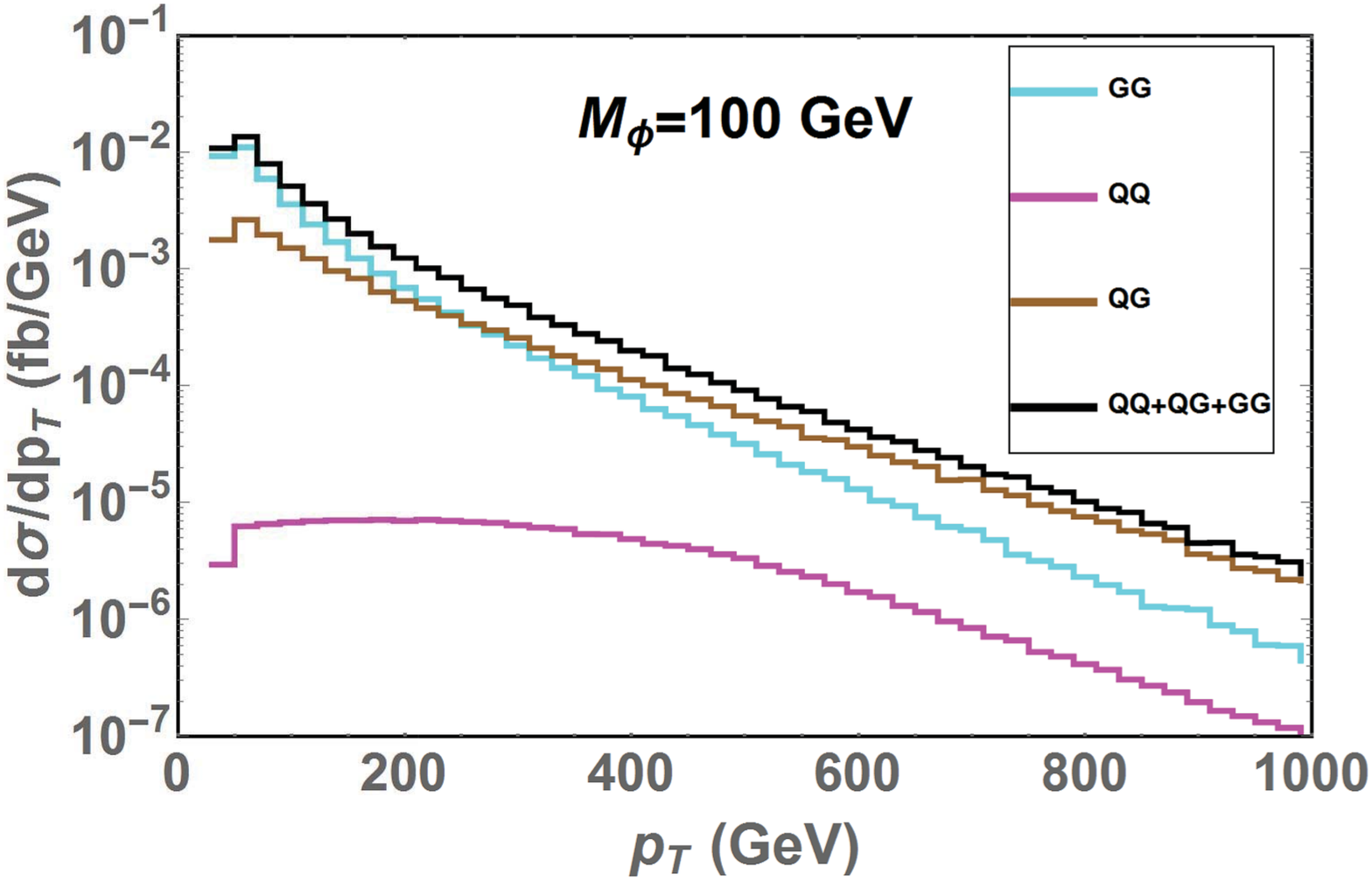}
\par\end{centering}

}\caption{Differential cross section with respect to $p_{T}^{j}$, for $p p\to \chi \chi^* j$
with $r=3$, $\lambda_d = 1$, $m_\chi = 1$~GeV, and $m_\phi$ as indicated.  The different lines show contributions from
the $gg$, $gq$, and $q\bar{q}$ initial state subprocesses, as well as their sum.\label{fig:InitialState} }
\end{figure*}

In Figure~\ref{fig:InitialState}, we show the differential cross section with respect to 
the jet transverse momentum, $p_{T}^{j}$.  At the parton level at leading order, this
quantity is the same as $\cancel{E}_{T}$.  We examine the relative importance of the subprocesses for a sample parameter point with $\lambda_d = 1$, a single species of mediator with $r=3$,
and a small dark matter mass\footnote{We choose a small dark matter mass $m_\chi = 1$~GeV
as an illustrative choice.  Results are typically insensitive to this particular choice for masses much less than the
cut on the mono-jet $p_T$.}
 $m_{\chi} = 1$~GeV.  We examine
two choices\footnote{Technically, $m_\phi = 10$~GeV is excluded by cosmological
considerations and the running of $\alpha_S$ \cite{Kaplan:2008pt}.  Nonetheless, it illustrates
the behavior for very low $m_\phi$ and is useful as a benchmark.} of $m_{\phi}=10$ and $100$~GeV.
We use the CTEQ6L1 parton distribution functions (PDFs) \cite{Nadolsky:2008zw}
and set the renormalization and factorization scales to $\mu = Q = H_T$. 
We observe that due a large gluon flux the $gg$ initial state dominates for smaller values of $p_{T}^{j}$. 
Note that for a given final state, the $gq$ flux dominates the $gg$ flux at sufficiently large $p_T$ scales. 
We also observe that at a higher $m_\phi$ value the $gq$ channel takes over the $gg$ channel at relatively smaller $p_T^j$ scale. 
On the other hand, the $q\bar{q}$ contribution remains small throughout due to the s-channel propagator suppression.

\subsection{Comparison with EFT \label{sec:EFT-applicability}}

\begin{figure*}[!t]
\subfloat[\label{fig:loopVSeft_8tev}$\sqrt{s}=$8 TeV]{\includegraphics[scale=0.195]{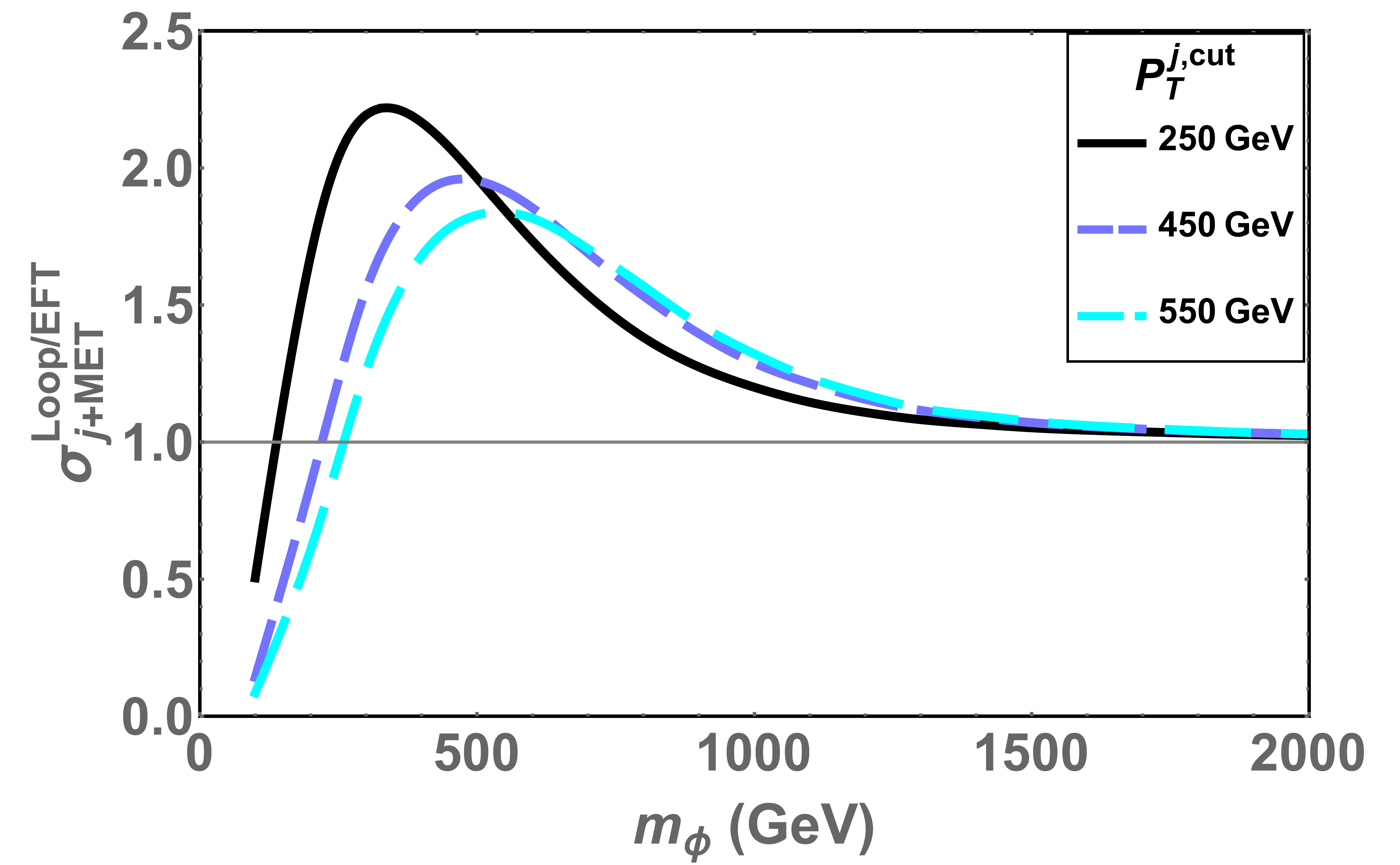}}
\hfill{}\subfloat[\label{fig:loopVSeft_13tev}$\sqrt{s}=$13 TeV]{\includegraphics[scale=0.195]{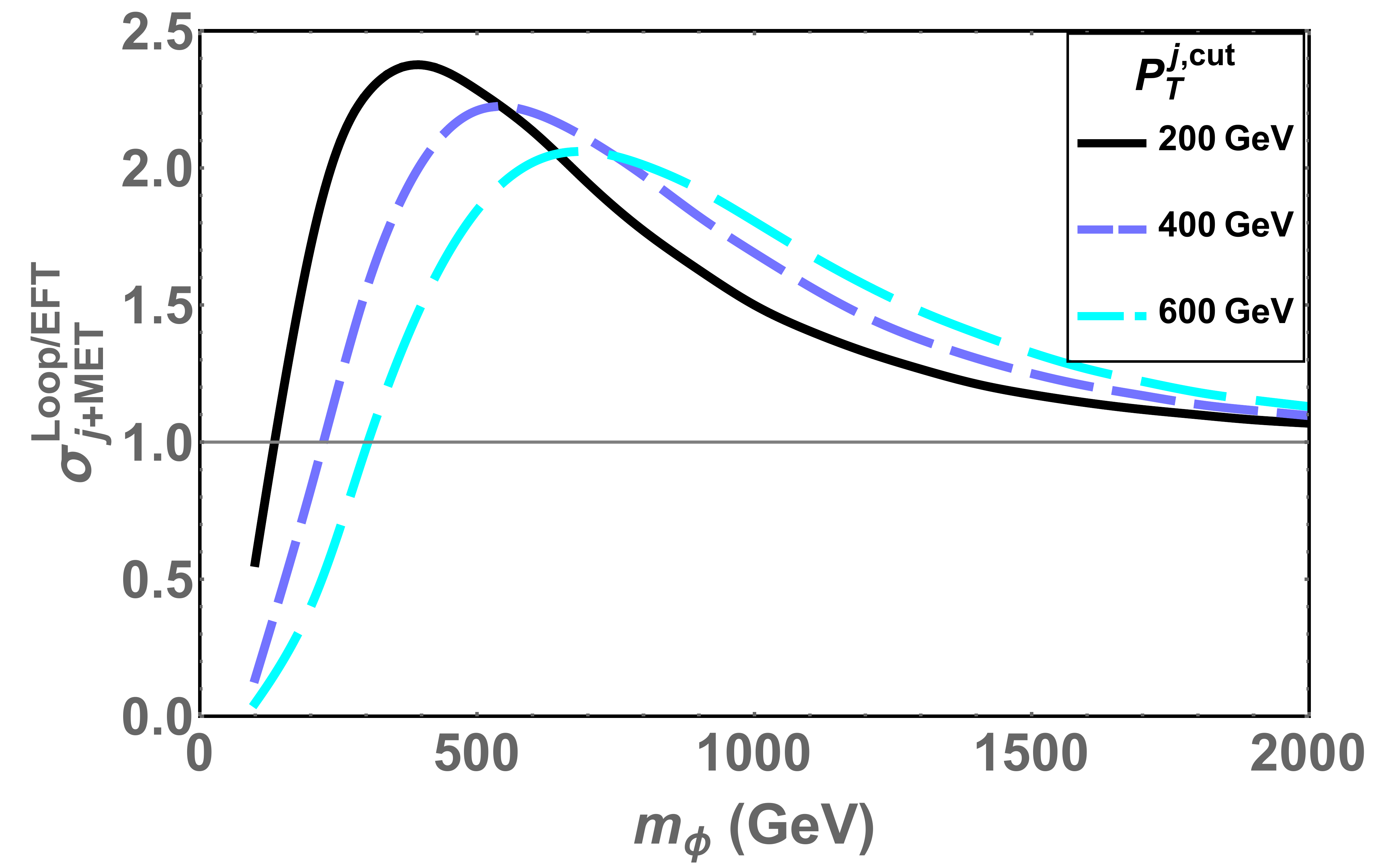}}
\caption{Ratio of the full calculation to the EFT approximation, as a function of $m_\phi$ (with other parameters fixed as described in the text), for three values of
the minimum $p_{T}^{j}$ and two choices of
center of mass energy, as indicated.}
\end{figure*}

In the limit $m_{\phi}\to\infty$, the full result is expected to flow to the one derived from the EFT, Eq.~(\ref{eq:eftmatching}).
In Figures~\ref{fig:loopVSeft_8tev} and \ref{fig:loopVSeft_13tev}, we show the ratio of the
full result to the EFT approximation for the sample parameter point defined above, as a function of
$m_\phi$, for $\sqrt{s} = 8$~TeV and $\sqrt{s} = 13$~TeV, respectively.
As expected, at small energy scales the 
EFT approximation over-estimates the cross section by a factor which
scales as $m_\phi^{-4}$.  It is interesting to note that the cross section calculated with loops becomes equal to that calculated in the EFT when 
the mediator mass is close to half the value of cut on jet transverse momentum ($m_\phi\sim p_{T}^{j}/2$).  At
scales comparable with the $p_{T}^{j}$ cut, EFT under estimates the cross
section by up to a factor of two. With a large cut on transverse missing
energy, the contributions from the resonant part of the $p_{T}$ distribution
in the case of a light scalar are removed and only the large $p_{T}$
region survives. In this region, EFT and full cross sections are comparable
and are both small.  In the asymptotic limit of large $m_\phi$, we find that the EFT
typically under-estimates the cross section by about $5\%$ at 8 TeV and by about $15\%$ at 13 TeV. 
Also, the merging of the three lines representing different $p_T^j$ cuts is faster at 8 TeV than at 13 TeV.
The fact that at 13 TeV the EFT under-estimates the cross section more than at 8 TeV is expected 
because at 13 TeV larger $p_T^j$ events are also accessible. This also suggests that for the higher energy runs of the future $pp$ collider, use of exact calculation would be preferable for this model.

\begin{table*}[!th]
\begin{centering}
\begin{tabular}{|c|c|c|c|c|c|c|c|c|c|c|c|}
\hline 
$p_{T}^{j}$~cut~ (GeV) & 150 & 200 & 250 & 300 & 350 & 400 & 450 & 500 & 550 & 600 & 700\tabularnewline
\hline 
\hline 
CMS-8 TeV $\sigma_{j + {\rm MET}}$~(fb) & ~~-~~ & ~~-~~ & ~~229~~ & ~~99~~ & ~~49~~ & ~~20~~ & ~~8~~ & ~~6~~ & ~~7~~ & ~~-~~  & ~~-~~ \tabularnewline
\hline 
\hline 
ATLAS-8 TeV $\sigma_{j + {\rm MET}}$~(fb) & ~~726~~ & ~~194~~ & ~~90~~ & ~~45~~ & ~~21~~ & ~~12~~ & ~~-~~ & ~~7~~ & ~~-~~ & ~~4~~ & ~~3~~ \tabularnewline
\hline
\hline 
ATLAS-13 TeV $\sigma_{j + {\rm MET}}$~(fb) & ~~-~~ & ~~-~~ & ~~553~~ & ~~308~~ & ~~196~~ & ~~153~~ & ~~-~~ & ~~61~~ & ~~-~~ & ~~23~~ & ~~19~~ \tabularnewline
\hline 
\end{tabular}
\par\end{centering}
\caption{\label{tab:monojet-upper-bounds}$95\%$ C.L. upper bounds
on BSM contributions to the mono-jet cross section obtained by the ALTAS and CMS experiments the LHC run I and II as a function of the cut on $p_{T}$ jet.}
\end{table*}

\begin{figure*}[!]
\subfloat[$r=3$]{\includegraphics[scale=0.19]{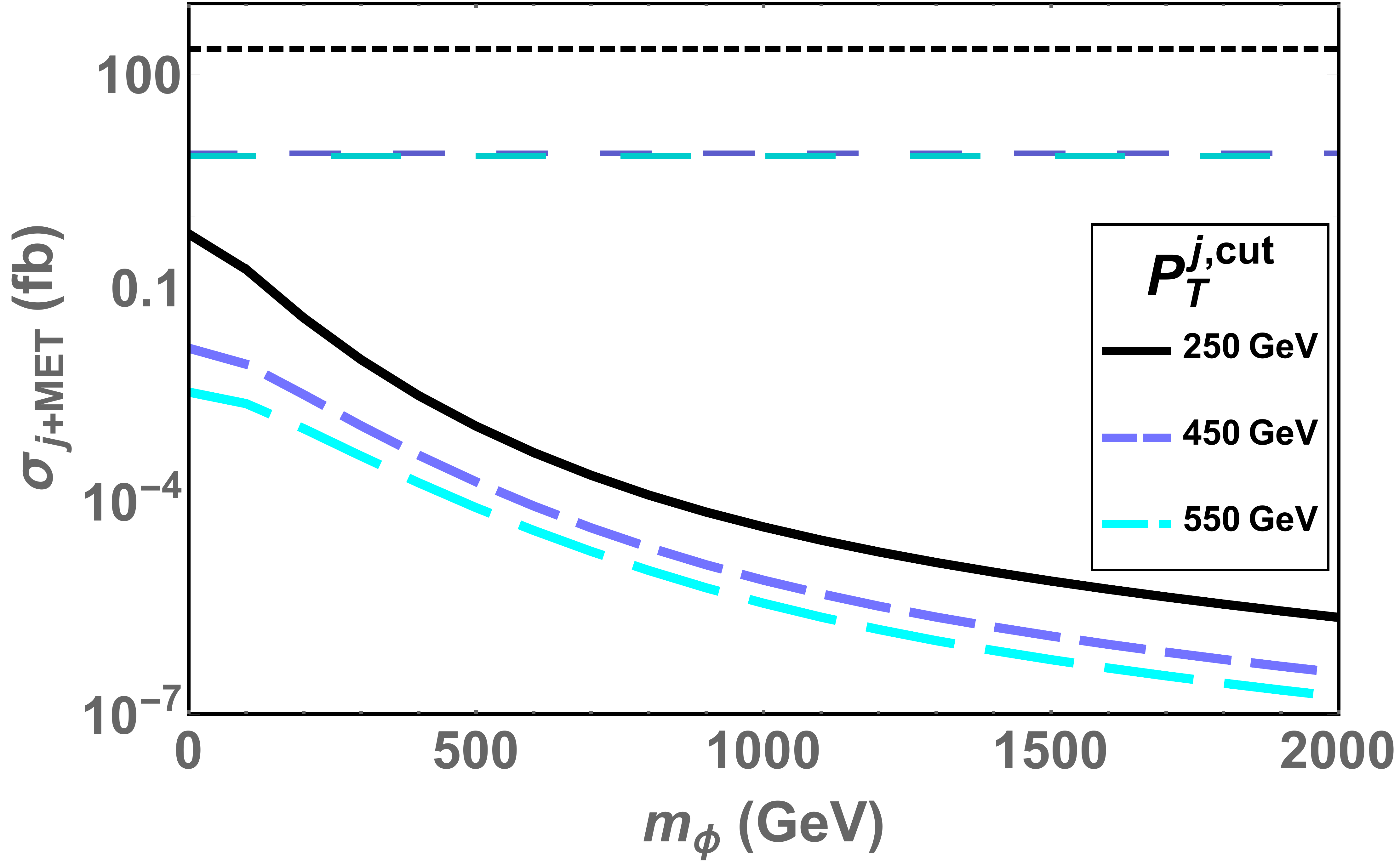}

}\hfill{}\subfloat[$r=15$]{\includegraphics[scale=0.19]{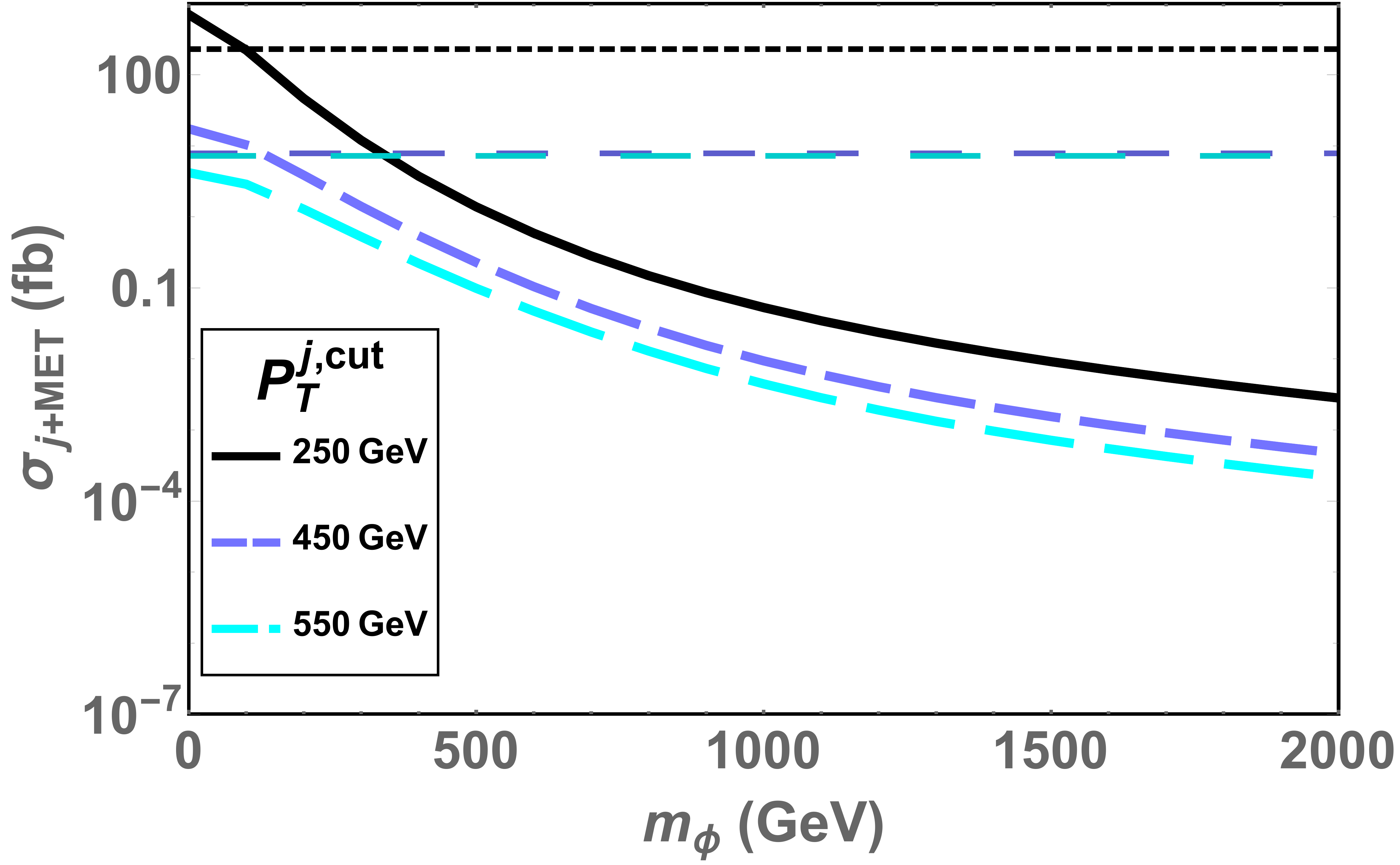}

}\caption{Mono-jet cross sections at 8 TeV as a function of the mediator mass $m_{\phi}$,
for three of the cuts on $p_{T}^{j}$ employed in the CMS analysis, as indicated. The horizontal lines
denote the $95\%$ C.L. bounds on the cross section, for the respective $p_{T}^{j}$ cuts. \label{fig:Loop-induced-monojet}}
\end{figure*}

\begin{figure*}[!t]

\subfloat[\label{fig:Loop-induced-monojet13TeV}$r=3$]{\includegraphics[scale=0.19]{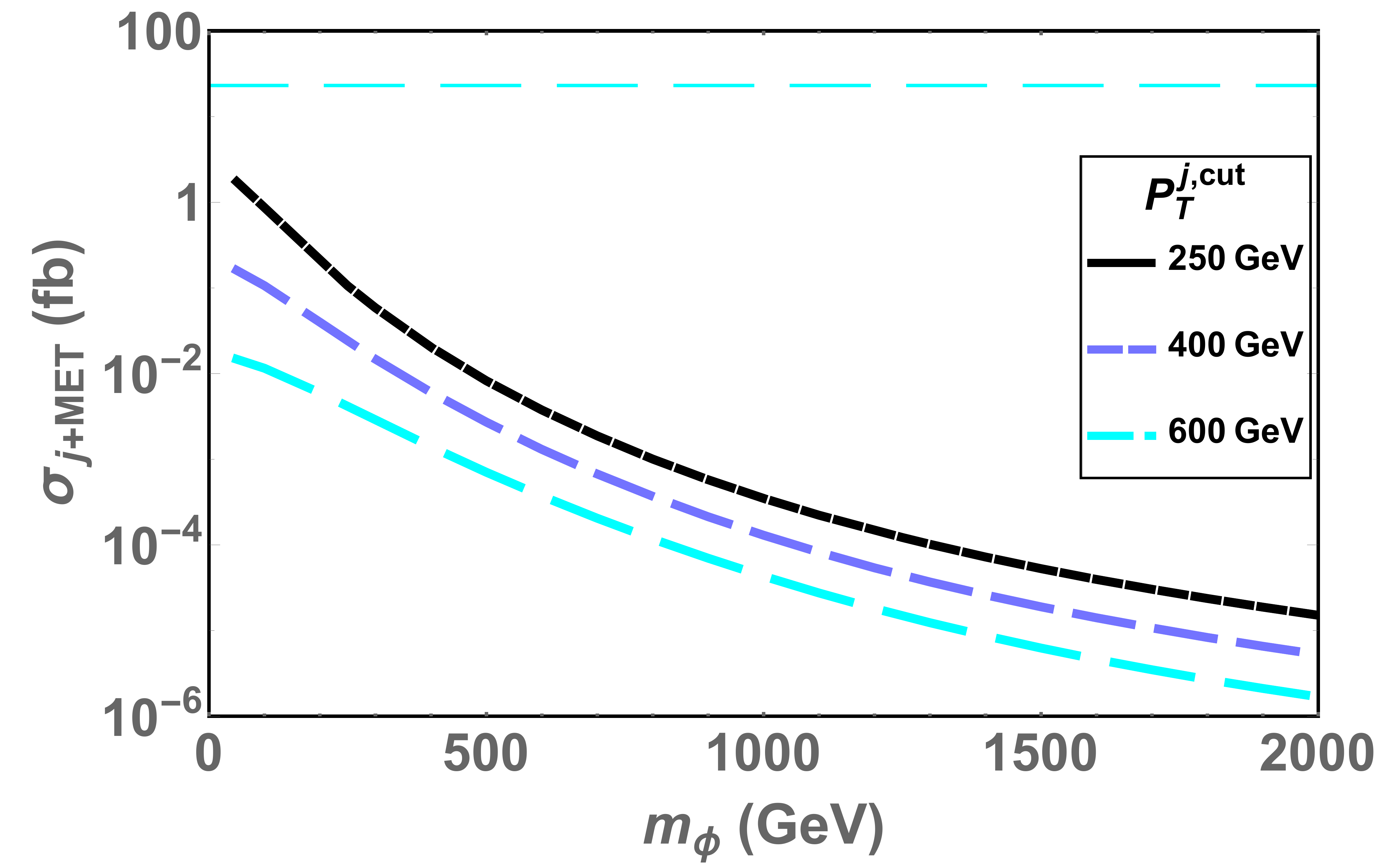}

}\hfill{}\subfloat[\label{fig:Loop-induced-monojet13TeV_1}$r=15$]{\includegraphics[scale=0.19]{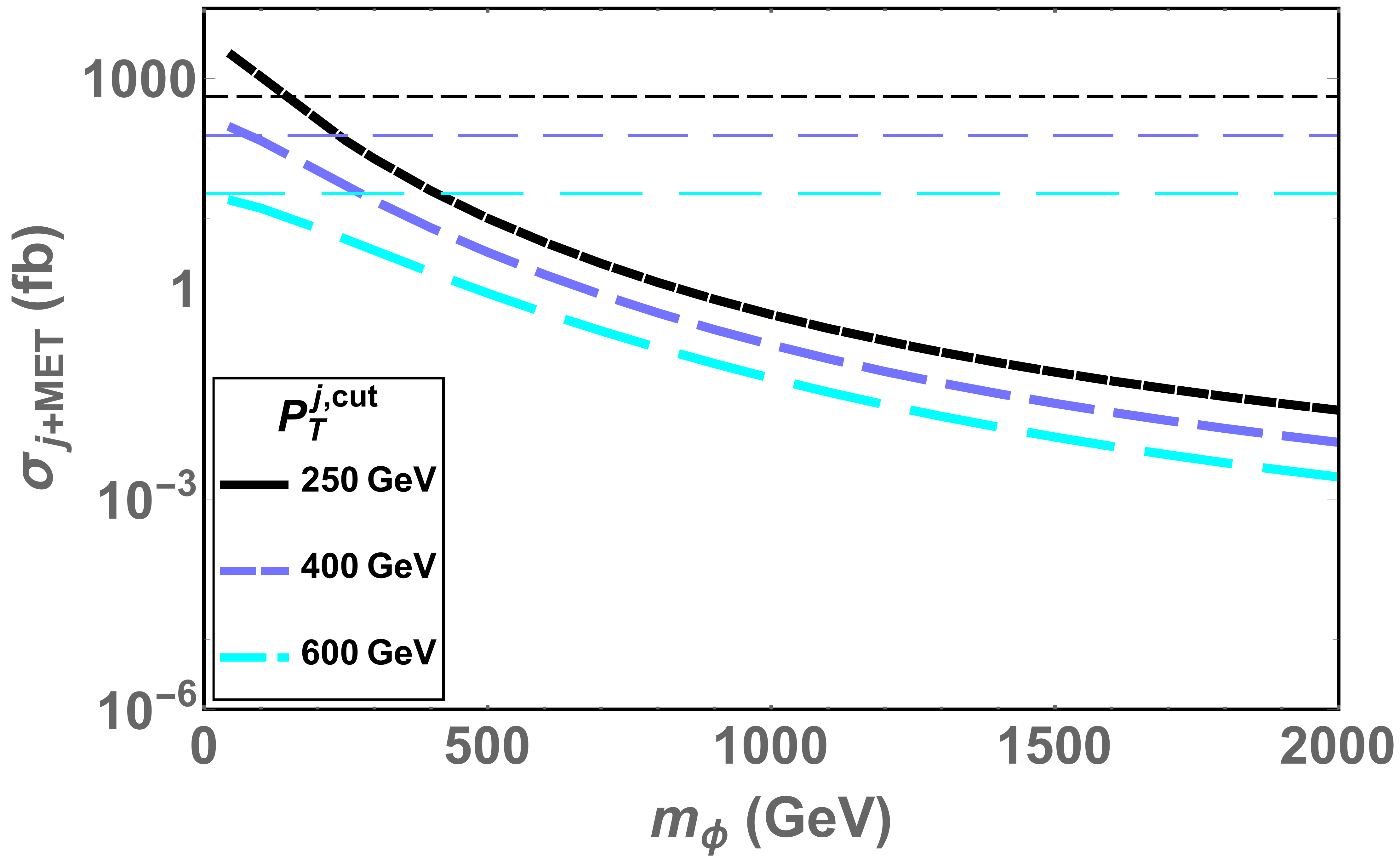}}
\caption{Mono-jet cross sections at 13 TeV as a function of the mediator mass $m_{\phi}$,
for three of the cuts on $p_{T}^{j}$ employed in the ATLAS analysis, as indicated. The horizontal lines denote the $95\%$ C.L. bounds on the cross section, for the respective $p_{T}^{j}$ cuts.}
\end{figure*}

\begin{figure*}[!th]
\subfloat[\label{fig:Reach-13TeV}$\sqrt{s}=$ 13 TeV]{\includegraphics[scale=0.19]{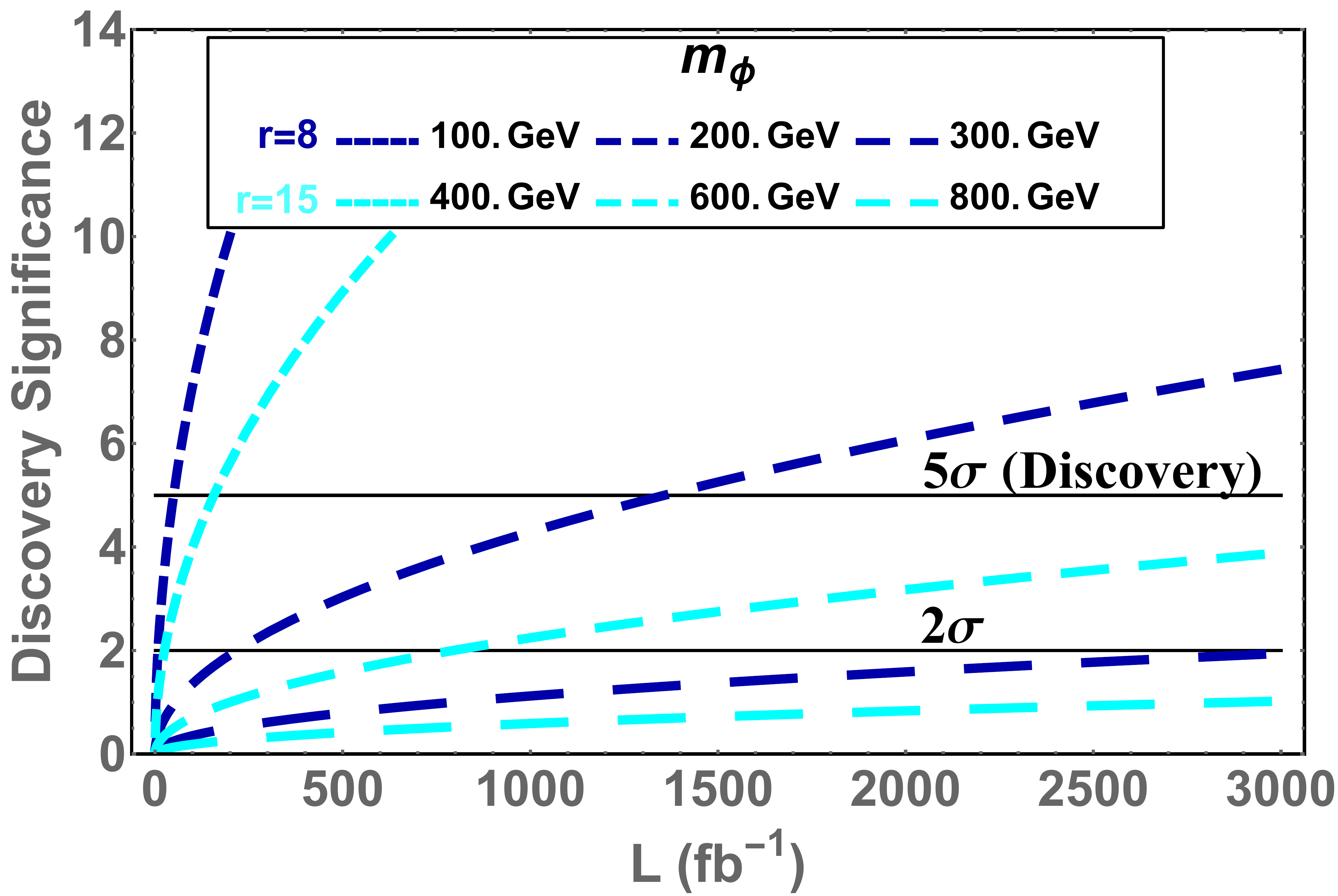}

}\hfill{}\subfloat[\label{fig:Reach-100TeV}$\sqrt{s}=$ 100 TeV]{\includegraphics[scale=0.19]{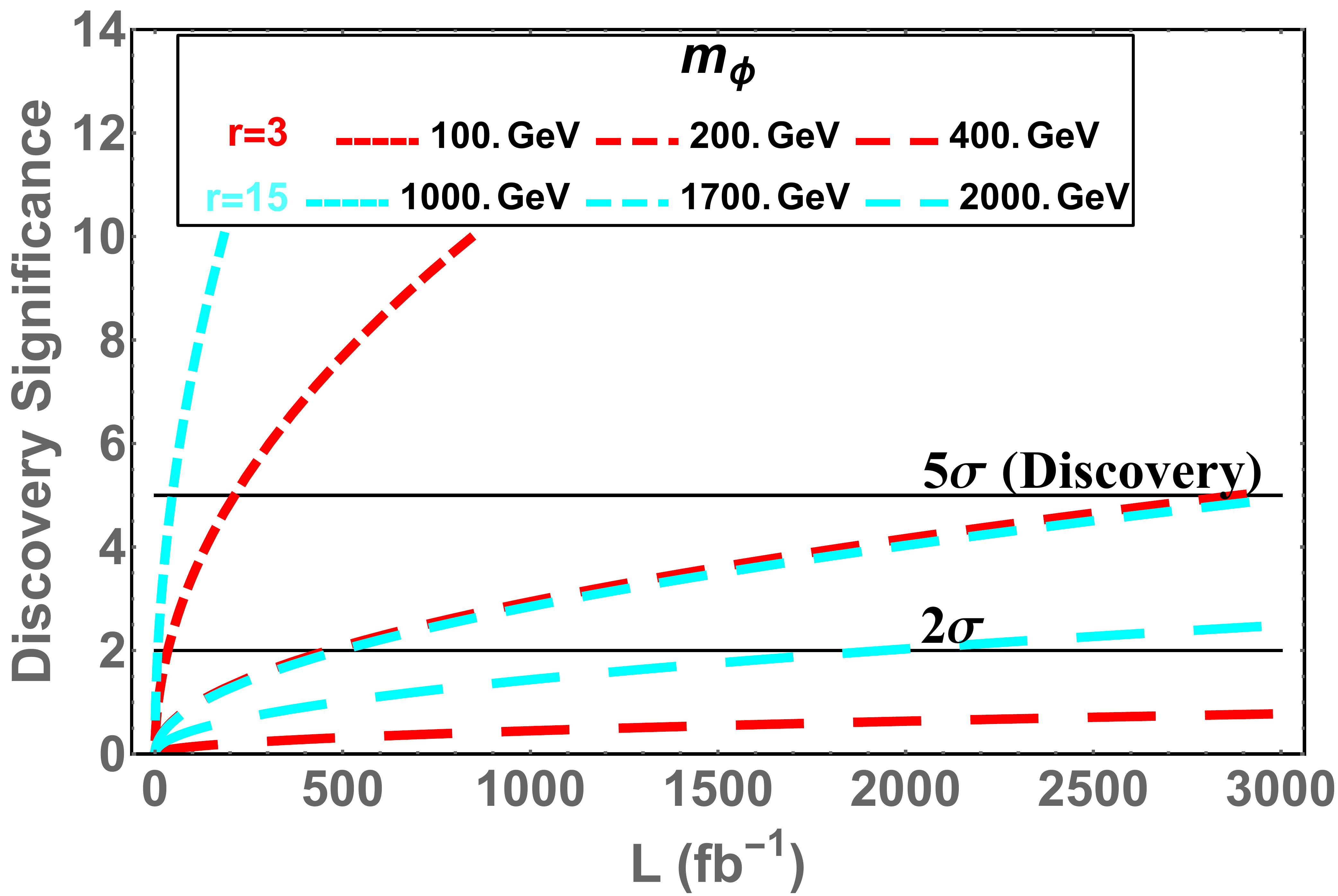}}
\caption{Significance ($S/\sqrt{S+B}$) of the mono-jet signal at the 13 TeV LHC and 100 TeV FCC
as a function of integrated luminosity, for mediators with $r=3$ (red), $r=8$ (dark blue), and $r=15$ (cyan), with a cut $p_{T}^{j} \ge 200$ GeV and masses as indicated on each figure. }
\end{figure*}

\section{Constraints and Future Prospects\label{sec:Results-and-projections}}

Next, we examine the minimal bounds from collider searches on the parameter space of a generic GSDM model. To this end, we begin with the bounds coming from dijet process and then perform a detailed analysis of the dark matter searches in the missing energy channel.

Production of a pair of jets is the most common process at the LHC. Strong constraints are obtained on a colored scalar interacting at the tree level with quarks \cite{Aaboud:2017yvp,CMS:2017xrr}. However, in the absence of couplings with the quarks, a colored scalar still contributes to this process albeit at one-loop level. This process is particularly important for BSM strongly interacting mediators as the only BSM parameters involved are the mass of the particle and its $SU(3)_C$ representation $r$. We use the framework described in the third technique in section \ref{sec:Calculation} and calculate the SM and BSM diagrams at next-to-leading order in $\alpha_{s}$. 

The NLO fixed order analysis of full dijet calculation in GSDM ($pp \to jj$) using {\tt MadGraph5\_aMC@NLO} indicates that an overall contribution of $\phi$
loops to the cross section is very small and negative. Using the cuts corresponding to the
ATLAS dijet search \cite{Aaboud:2017yvp}, we find that the effects are of the order of 0.2\% and 1.4\% for the color triplet and color octet cases respectively for a very light mediator mass, $m_\phi$ (~1GeV). The corresponding effects with CMS cuts \cite{CMS:2017xrr} are at the level of 1\% and 6.7\% respectively.
Therefore, any meaningful constraints on $m_\phi$ from the available dijet data is
expected only for very large dimensional color representations. A more detailed
investigation of the calculation is ongoing and will be reported in a future work \cite{Godbole:2018xyz}.

In the remainder of this study, we focus on the mono-jet search as the observable most intimately connected to the dark matter itself, and also the most model-independent probe available.  Searches for the mediators based on their
production and decay into jets are rather model-dependent.  For example,
the minimally flavor violating (MFV) color triplet scalar analyzed in Ref. \cite{Godbole:2015gma} must have
$m_{\phi_3} \ge 350$ GeV to satisfy 4-jet searches \cite{Khachatryan:2014lpa},
with only mild dependence on its coupling to quarks,
whereas a color octet scalar receives much stronger bounds of $m_{\phi_8}\ge3.1$ TeV from
searches for resonances in dijet production \cite{Khachatryan:2015dcf}.

\subsection{Constraints from 8 TeV}

We explore the range of parameter space probed by the run I search for mono-jets
\cite{Khachatryan:2014rra,Aad:2015zva}.  These bounds are presented as a $95\%$ C.L. upper bound on 
the cross section for any beyond-the-SM contribution to the mono-jet signature ($\sigma_{j + {\rm MET}}$)
as a function of the
cut on $p_{T}^{j}$, and the experimental results are summarized in the first and second row of Table~\ref{tab:monojet-upper-bounds}.
The pseudo-rapidity of the leading jet is further required to satisfy $|\eta^{j}|<2.4$ in the experimental analysis of CMS and $|\eta^{j}|<2.0$ for ATLAS.

We apply the experimental selection to our full calculation of the mono-jet cross section, continuing to examine the case of
$\lambda_d = 1$ and light dark matter, $m_{\chi}=1$ GeV.  We choose three representative $p_{T}^{j}$ cuts from the
CMS analysis, and show the resulting cross section after cuts in Figure~\ref{fig:Loop-induced-monojet}, for
two choices of mediator representation, $r=3$ and $r=15$.  Also shown are the corresponding limits on the cross section
for the respective choice of $p_{T}^{j}$ cut.  Comparing the two, we find that the color triplet mediator is
completely unconstrained by the current mono-jet bounds, whereas the $r=15$ representation is subject to very mild
bounds of order $m_\phi \gtrsim 158$ GeV, obtained from the ATLAS run-I data with a $p_{T}^{j} \geq 350$~GeV.

\subsection{Constraints from 13 TeV}

In Figures~\ref{fig:Loop-induced-monojet13TeV} and \ref{fig:Loop-induced-monojet13TeV_1}, we show the mono-jet cross section at LHC run-II
as a function of $m_\phi$, for $\lambda_d=1$, and $m_\chi = 1$~GeV  with $r=3$ and
$r=15$, respectively, for a few representative choices of the $p_{T}^{j}$ cuts from the ATLAS run-II analysis \cite{Aaboud:2016tnv}. The limits obtained on the value of $m_\phi$ from the run-II analysis with 3.2 fb$^{-1}$ of data are weaker than the corresponding run-I results.

\subsection{Future Prospects}

We examine the prospects for future colliders to probe the parameter space of GSDM through searches for the mono-jet process. To assess the reach of these colliders to discover GSDM for different values of $m_\phi$, we compute the primary
(irreducible) SM background to the mono-jet process from $Z+j$ production, where the $Z$ boson decays into neutrinos.
We compute this background at leading order for the 13 TeV LHC and for the proposed 100 TeV FCC using Madgraph,
subject to the cuts on the mono-jet: $|\eta_{j}|<2.4$, and a modest cut of $p_{T}^{j}>200$ GeV.  
We assume that, as was true for the LHC run I analysis, the real background from $Z+j$ dominates over the
fake contribution from mis-measured QCD jets.
In Figures~ \ref{fig:Reach-13TeV}
and \ref{fig:Reach-100TeV} we present the significance, defined as $S/\sqrt{S+B} \simeq S/\sqrt{B}$ as a function of the
integrated luminosity at each accelerator.

We find that with 3 ab$^{-1}$ of luminosity, the 13 TeV LHC can discover (at $5\sigma$) evidence for a color octet mediator
whose mass is slightly above 200 GeV.  A 15 of color reaches $5\sigma$ discovery for masses around 500 GeV.  Obviously,
a much larger range of parameter space can be explored for higher dimensional
representations, even with lower luminosities.
At the FCC, the reach for a color triplet
scalar in the mono-jet channel reaches the level of discovery
for masses up to $m_{\phi}\sim200$ GeV.  A much
larger range of parameter space can be explored for higher dimensional
representations: for $r=15$, masses up to 1.7 TeV can be probed with 3 ab$^{-1}$.

\section{Summary\label{sec:Summary}}

A scalar gauge singlet dark matter particle allows for the possibility of a renormalizable connection to the SM
via a quartic interaction with a scalar particle which is itself charged under the Standard Model.  If this scalar
is colored, the result is a loop level connection predominantly to the SM gluons, resulting in
phenomenology at a hadron collider with unique features.  
In this article, we have computed the full one-loop contribution to the mono-jet process
(focused on the regime of light dark matter, $m_\chi \sim 1$~GeV)
at the LHC and a proposed 100 TeV accelerator.  The full one-loop treatment is crucial to probe the regime of light
mediator masses and asses the viable parameter space, where the EFT treatment does not apply.
We find that low dimensional SU(3) representation mediator is very difficult to probe at the LHC; even with large
datasets the bounds are not competitive with the generic ones from LEP II.  However, larger representations can
be discovered up to masses of several hundred GeV with large data sets.  A future 100 TeV collider
can probe a much larger parameter space, reaching TeV scale masses.

\section*{Acknowledgments}

RMG wishes to acknowledge support from the Department
of Science and Technology, India under Grant No. SR/S2/JCB-64/2007,
 under the J.C. Bose Fellowship scheme.The research of TMPT is supported in part by NSF grant PHY-1316792 and by the University of California, Irvine through a Chancellor's
Fellowship. AS would like to thank Iaonnis Tsinikos, Manoj K. Mandal and Olivier Mattelaer for their help with MadGraph5 aMC@NLO package. 

\bibliography{monojet}

\end{document}